\documentclass[twocolumn,showpacs,preprintnumbers,amsmath,amssymb,superscriptaddress,prb]{revtex4}
\usepackage{graphicx,pgfplots}
\usepackage{dcolumn}
\usepackage{bm}

\begin{document}
\preprint{APS/123-QED}

\title{Point-contact tunneling spectroscopy measurement of Cu$_x$TiSe$_2$: disorder-enhanced Coulomb effects}

\author{Katherine Luna}
\affiliation{Department of Physics, Stanford University, Stanford CA 94305-4045, USA}\author{Phillip M. Wu}
\affiliation{Department of Physics, Stanford University, Stanford CA 94305-4045, USA}
\author{Justin S. Chen}
\affiliation{Department of Physics and Astronomy, Rice University, Houston, TX, 77005, USA}
\author{Emilia Morosan}
\affiliation{Department of Physics and Astronomy, Rice University, Houston, TX, 77005, USA}
\author{Malcolm R. Beasley}
\affiliation{Department of Physics, Stanford University, Stanford CA 94305-4045, USA} 
\date{\today}

\begin{abstract}
We performed point-contact spectroscopy tunneling measurements on Cu$_x$TiSe$_2$ bulk with $x=0.02$ and $0.06$ at temperatures ranging from $T=4-40$ K and observe a suppression in the density of states around zero-bias that we attribute to enhanced Coulomb interactions due to disorder.  We find that the correlation gap associated with this suppression is related to the zero-temperature resistivity.  We use our results to estimate the disorder-free transition temperature and find that the clean limit $T_{c0}$ is close to the experimentally observed $T_c$.
\end{abstract}

\maketitle

Copper intercalated titanium diselenide (Cu$_x$TiSe$_2$) is a fascinating system offering a unique opportunity to study the interplay of two collective phenomena, namely superconductivity and charge density waves (CDW). \cite{Morosan06,Li07_2,Qian07, Wezel10} The parent compound TiSe$_2$ has been classified as either a CDW semi-metal or excitonic insulator, \cite{Hellmann12} and upon the addition of Cu, superconductivity arises with a maximum transition temperature $T_c=4.15$ K near $x\approx0.08$.  Several experiments have already been conducted to probe the relationship between these two states.  Photoemission studies have shown that the CDW order parameter microscopically competes with superconductivity in the same band.\cite{Qian07}  In addition, previous work suggests that the CDW is suppressed by increasing the chemical potential, while superconductivity is enhanced by the increasing density of states (DOS).\cite{Zhao07}

With an increase in the chemical potential combined with the observation of the rapidly varying DOS near the Fermi energy, $E_F$, the question of the impact of disorder on this system is relevant.\cite{Jeong07}  Band structure calculations suggest that disorder may play an important role in moderating the large DOS. The effects of disorder then would allow for increased orbital hybridization, effectively increasing electron-phonon coupling $\lambda$ despite a reduced total DOS.  In this paper we report our findings of point-contact spectroscopy (PCS) measurements on Cu$_x$TiSe$_2$ and analyze our results in the context of disorder. We find that disorder-enhanced Coulomb interactions are present in the superconducting and non-superconducting crystals, suggesting disorder does play a role in these materials. 

\begin{figure}[t!]
  \centering
  \includegraphics[width=0.5\textwidth]{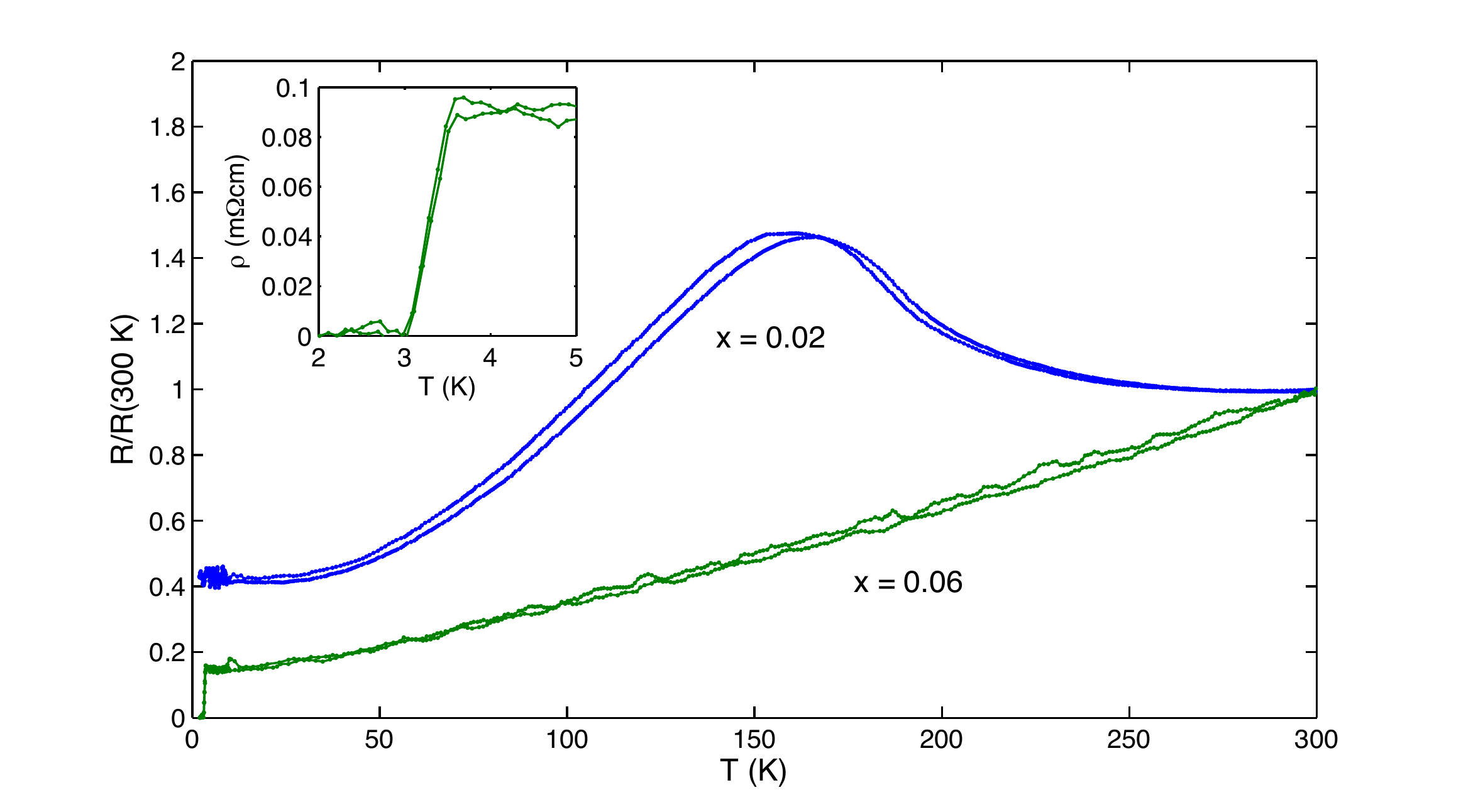}
  \caption{Residual resistance ratio versus temperature for $x=0.02$ (blue curve) and $x=0.06$ (green curve) samples. The inset shows the superconducting transition with $T_c=3$ K for the $x=0.06$ sample. The hump at roughly 150 K in the $x=0.02$ sample is a signature of the CDW.}
\label{fig:rvt}
\end{figure}

\begin{figure*}[t!]
  \centering
  \includegraphics[width=0.9\textwidth]{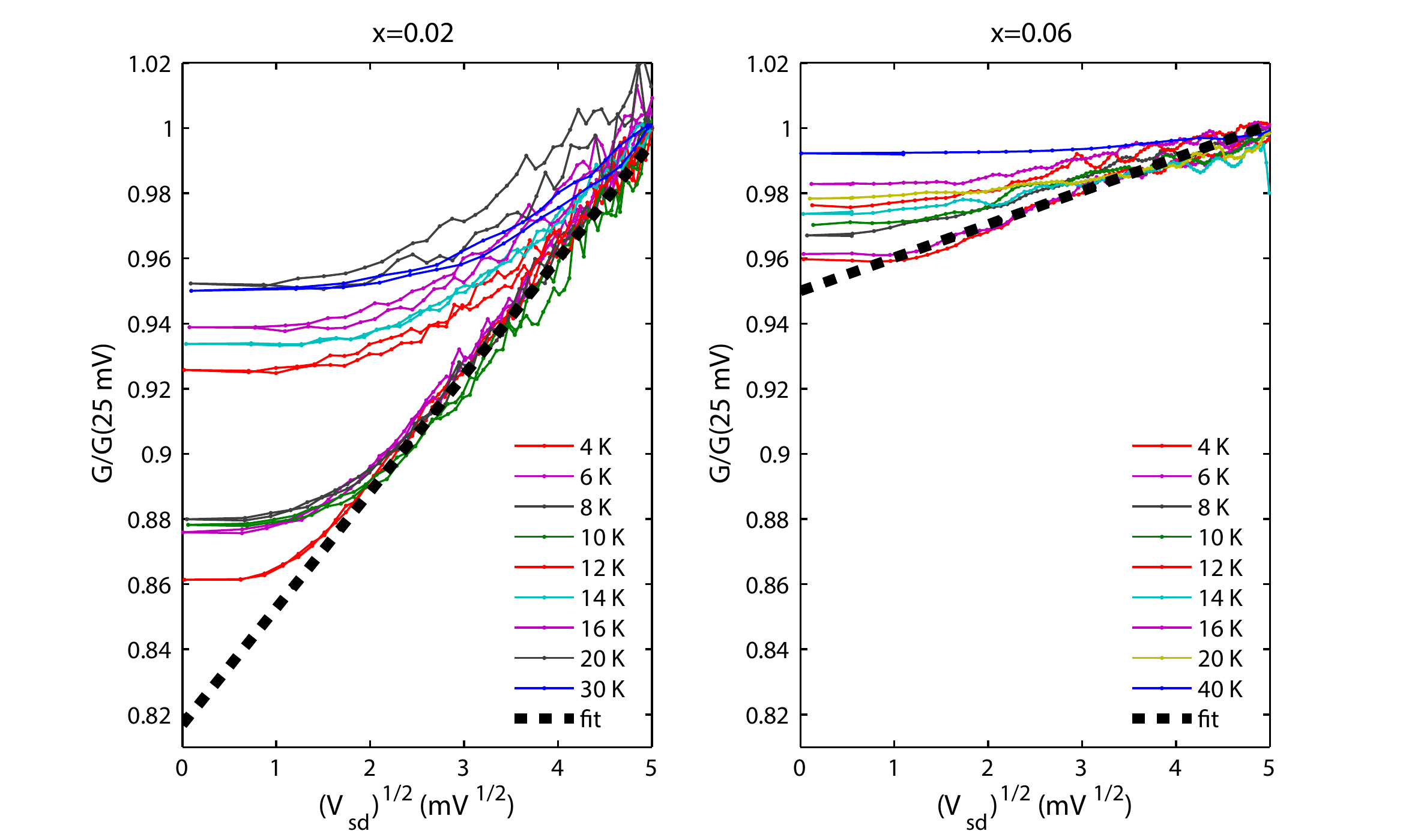}
  \caption{Normalized differential conductance G versus square root of the source drain voltage $V_{sd}$ in units of $(mV)^{1/2}$ for Cu$_x$TiSe$_2$ with (a) $x=0.02$ and (b) $x=0.06$. The black dashed lines are fits used to extract the normalized zero temperature DOS and correlation gap.}
\label{fig:CuTiSe2_PCS}
\end{figure*}

The samples were grown by the method described previously \cite{Morosan06, Morosan07}.  Four-terminal resistance versus temperature measurements for the crystals studied here are shown in Fig. \ref{fig:rvt}. The $x=0.02$ sample (upper curve),  with a resistivity $\sim$ 1-2 $m\Omega$cm at low temperature, is not superconducting, while the resistivity of the sample with $x=0.06$ is on the order of 0.1 $m\Omega$cm just above the superconducting transition at $T_c=3$ K. The $x=0.02$ sample displays a pronounced bump near 150 K, a signature of the CDW. This feature is not evident in resistivity curves for the superconducting sample, consistent with the suppression of the CDW below $0.4$ K when $x> 0.04$.\cite{Morosan06} To further understand the gap structure, we performed PCS measurements on these two samples using an oxidized Aluminum tip of diameter $0.5$ mm. Point contact measurements were performed at temperatures ranging from $4-40$ K. The normalized differential conductance, $G/G(25 mV)$, shown in Fig. \ref{fig:CuTiSe2_PCS} allows for a direct measure of the normalized DOS $N(E)$ \cite{Luna13, Wu2014}. For the superconducting sample, we are essentially probing the behavior in the normal state in this temperature range. The junctions were prepared by cleaving the sample in air, and then bringing the sample in contact with the tip at room temperature.  The apparatus was then inserted into a flow cryostat for measurements.  Previous studies using these tips demonstrate that they provide a tunneling contact.\cite{Luna13, Wu2014}

We observe a suppression in the DOS near zero-bias as indicated by the cusps in Fig. \ref{fig:CuTiSe2_PCS} in the normal state. These cusps are reminiscent of those observed in amorphous Nb-Si alloys,\cite{Hertel83} which is one of the classic cases of a disorder driven metal-insulator transition, and more recently such cusps were observed in BaPb$_{1-x}$Bi$_x$O$_3$ and alkali-doped tungsten bronzes.\cite{Luna13, Wu2014} In disordered metals, the reduction of $N(E)$, due to disorder-enhanced Coulomb interactions is well established.  In three dimensions theory predicts that 
\begin{equation}
N(E) = N(0)[1+(E/\Delta)^{1/2}],
\label{eq:DOS}
\end{equation}
 where $N(0)$ is the normalized DOS at zero temperature, $\Delta$ is the correlation gap and $E=V_{sd}$ is the source drain voltage.\cite{Altshuler79}  

To aid in the analysis, in Fig. \ref{fig:CuTiSe2_PCS} we show the normalized differential conductance versus the square root of the source drain voltage in units of $(mV)^{1/2}$ for Cu$_x$TiSe$_2$ with (a) $x=0.02$ and (b) $x=0.06$. Clearly, our data follows the energy dependence in Eq. \ref{eq:DOS} relatively well. The black dashed lines in both panels are fits to the data at the lowest temperature measured.  From this fit, we determine both the correlation gap $\Delta$, which corresponds to the inverse slope of the line and the zero-temperature reduction in the normalized DOS at zero-bias, N(0), corresponding to the zero voltage intercept of the dashed line. For $x=0.02$, we find $N(0)=0.82$ and $\Delta=0.52$ eV. For $x=0.06$, we find $N(0)=0.95$ and $\Delta=8.6$ eV.

We note that the superconducting sample, $x=0.06$ has a larger $N(0)=0.95$ than the non-superconducting sample. This trend is consistent with the fact that Cu doping should increase the DOS while also raising the chemical potential. Also, the one order of magnitude larger correlation gap of $\Delta=8.6$ eV in the superconducting sample $x=0.06$ suggests a greater role of disorder-induced Coulomb interactions.


\begin{figure}[t!]
  \centering
  \includegraphics[width=0.4\textwidth]{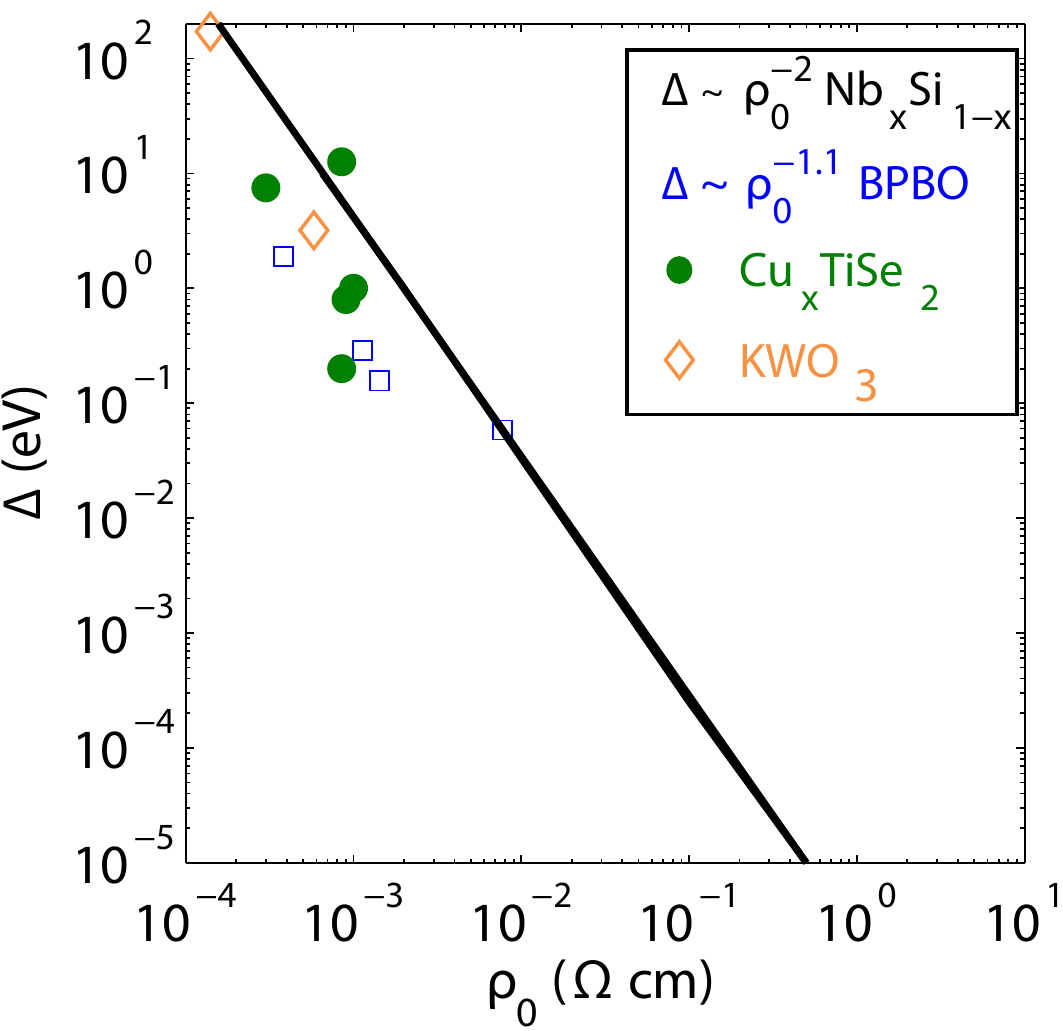}
  \caption{Correlation gap as a function of the zero-temperature resistivity for Nb$_x$Si$_{1-x}$ (black line), BaPb$_{1-x}$Bi$_x$O$_3$ (blue squares), K$_{0.33}$WO$_3$ thin films with different thicknesses (orange diamonds) and Cu$_x$TiSe$_2$ (green circles) for Cu concentration $x=0.02$ and $0.06$.}
\label{fig:gap-doping}
\end{figure}

For comparison, in Fig. \ref{fig:gap-doping}, we plot the correlation gap, $\Delta$, as a function of the zero-temperature resistivity, $\rho_0$, of various samples and concentrations.  The solid black line is the relationship found for Nb$_x$Si$_{1-x}$.\cite{Hertel83}  The magnitude of the correlation gap in Cu$_x$TiSe$_2$ is roughly consistent with the other materials, suggesting universality of disordered enhanced Coulomb interactions in these distinct classes of metals. 



Finally we turn to the question of the effect of disorder on $T_c$ and apply the same formalism as was used in \cite{Luna13, Wu2014}. Briefly, from the work of Belitz,\cite{Belitz89} we have a modified McMillan equation for $T_c$ valid for strong coupling and relatively strong disorder.

\begin{equation}
T_c = \frac{\Theta_{D}}{1.45}\textrm{exp}\left[\frac{-1.04(1+\tilde{\lambda}+ Y')}{\tilde{\lambda} - \tilde{\mu}^*[1+0.62\tilde{\lambda}/(1+Y')]}\right].
\label{TcBelitz}
\end{equation}

Here $\Theta_D$ is the Debye temperature, $\tilde{\lambda}$ is the disorder dependent electron-phonon coupling, and $\tilde{\mu^*}$ is the disorder dependent Coulomb pseudopotential. Conveniently, the disorder is parameterized by the fractional reduction of the DOS at the Fermi energy.
 
\begin{equation}
Y' = N(E_F)/N(0)-1
\end{equation}
where $N(E_F)$ is the DOS at the Fermi level.
 
$Y'$ enters in the equation for the reduction of $T_c$ both explicitly as shown in Eq. \ref{eq:DOS}  and implicitly through $\tilde{\lambda}(Y')>\lambda$ and $\tilde{\mu}^*(Y')>\mu^*$.   

\begin{figure}[t!]
  \centering
  \includegraphics[width=0.4\textwidth]{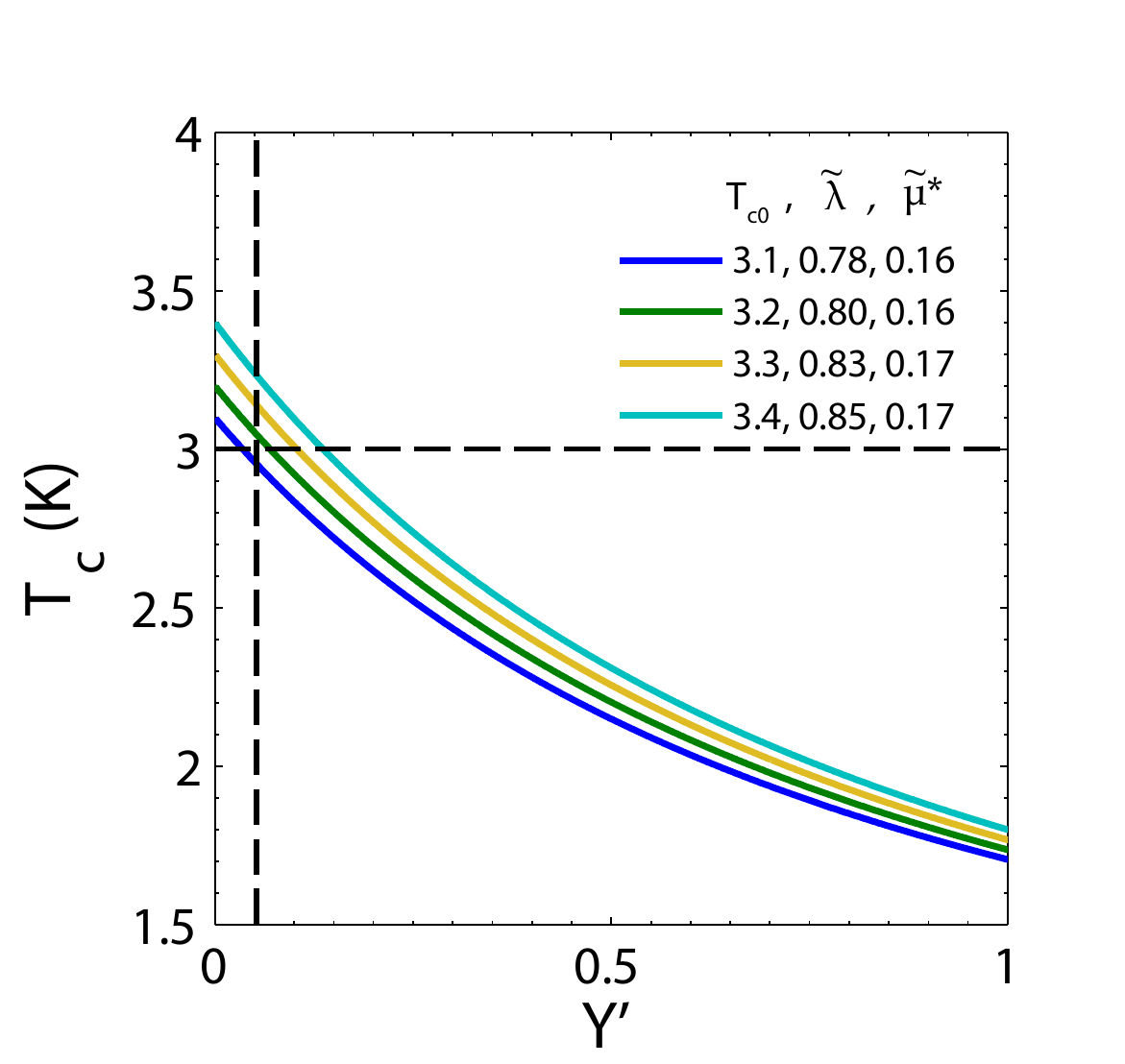}
  \caption{Calculated $T_c$ as a function of the disorder parameter $Y'$ for different values of $\tilde{\mu^*}$ and $\tilde{\lambda}$.  From the intersection of the measured $T_c$  and $Y'$ for $x\approx0.06$, $T_{c0}$ where $Y'=0$ can be backtracked.}
\label{fig:TcSuppression_Belitz_Cu06TiSe2_ai}
\end{figure}

We can determine the clean limit transition temperature without disorder, $T_{c0}$, by first plotting an array of curves for the variation of $T_c$ with the disorder parameter $Y'$, as shown in Fig. \ref{fig:TcSuppression_Belitz_Cu06TiSe2_ai}. We can then triangulate the particular curve of interest, as we know the measured $T_c$ and can estimate $Y'$ using the DOS from PCS. Tracing back the selected curve, corresponding to the intersection point, to when $Y'=0$ yields for $x\approx0.06$,  $(T_c, T_{c0})=(3 \mathrm{K}, 3.2 \mathrm{K})$, $(\lambda,\tilde{\lambda})=(0.76,0.8)$, $(\mu^*,\tilde{\mu^*})=(0.15,0.16)$.

Using this formalism, we find that the observed $T_c=3$ K is not significantly different from the clean limit $T_{c0}=3.2$ K, in contrast to BPBO or KWO$_3$.\cite{Luna13,Wu2014} There are some differences between these materials which might account for the possible effects disorder has on $T_c$. In Cu$_x$TiSe$_2$, the CDW partially gaps the Fermi surface\cite{Zhao07} whereas in BPBO, for example, the negative-U CDW insulator affects the entire Fermi surface, so the effect of disorder on $T_c$ in BPBO is more pronounced.\cite{Franchini09, Franchini10}

In summary, we performed PCS on Cu$_x$TiSe$_2$ and found a suppression in the DOS. Our results suggest that disorder-enhanced Coulomb interactions should be taken into account in this system. We found that the correlation gap associated with this suppression is correlated with the zero-temperature resistivity, suggestive of some universal scaling result across a number of materials. Finally, we address the question of how disorder impacts $T_c$. We find that the clean limit $T_{c0}$ is close to the experimentally observed $T_c$, suggesting that the effects of disorder on $T_c$ are not very pronounced.

\begin{acknowledgments}
This work supported by Air Force Office of Scientific Research MURI grant FA9550-09-1-0583-P00006.  J.C. and E.M. acknowledge support from the DOD PECASE award.
\end{acknowledgments}


\newpage
\bibliography{CuTiSe2bib}

\end{document}